\documentclass[nofootinbib,preprint,aps]{revtex4}
\usepackage{graphicx}

\begin{document}
\title{
Gravity induced neutrino$-$antineutrino oscillation: CPT and lepton number non-conservation 
under gravity
}

\author{Banibrata Mukhopadhyay\footnote{bm@physics.iisc.ernet.in; bmukhopa@cfa.harvard.edu}}

\affiliation {1. Department of Physics, Indian Institute of Science, Bangalore-560012,
India\\2. Theory Division, Harvard-Smithsonian Center for Astrophysics, 60 Garden Street, 
MS-51, Cambridge, MA 02138, USA
}


\def\lsim{\lower.5ex\hbox{$\; \buildrel < \over \sim \;$}}
\def\gsim{\lower.5ex\hbox{$\; \buildrel > \over \sim \;$}}

\def\ch{\lower-0.55ex\hbox{--}\kern-0.55em{\lower0.15ex\hbox{$h$}}}
\def\lh{\lower-0.55ex\hbox{--}\kern-0.55em{\lower0.15ex\hbox{$\lambda$}}}
\def\n{\nonumber}

\begin{abstract}

We introduce a new effect in the neutrino oscillation phase which shows the 
neutrino$-$antineutrino oscillation is possible under gravity even if the rest masses 
of the corresponding eigenstates are same. This is due to CPT violation
and possible to demonstrate if the
neutrino mass eigenstates are expressed as a combination of neutrino and antineutrino
eigenstates, as of the neutral kaon system, 
with the plausible breaking of lepton number conservation. 
For Majorana neutrinos, this oscillation is expected to affect
significantly the inner edge of neutrino dominated accretion disks around a compact object
by influencing the neutrino sphere which controls the accretion dynamics, and then the 
related type-II supernova evolution and the r-process nucleosynthesis.
On the other hand, in early universe, in presence of various lepton number violating processes,
this oscillation, we argue, might lead to neutrino asymmetry which
resulted baryogenesis 
from the $B-L$ symmetry by electro-weak sphaleron processes.

\end{abstract}

\maketitle

\section{Introduction}

It is well known that the discrepancy
between various observed and expected number of neutrino
flavors, e.g. $\nu_e$, $\nu_\mu$ and $\nu_\tau$, occurs because they transform from 
one type of flavor to another, namely flavor oscillation, which conserves lepton number. 
Therefore, if a detector is built and set to
detect for one type of neutrinos, while that type of neutrinos
are coming, some of them convert to another type which is
unsuited to the detector and thus results a shortfall. 
This anomalies have been found in the cases of solar neutrino,
atmospheric neutrino and the neutrino experiments at accelerators or reactors. 
This is also known that the neutrino flavor states consist of superposition
of different mass eigenstates which evolve differently with time and thus flavor states transform from
one to another.  
However, if CPT
is violated, even for a particular flavor the neutrino and the
anti-neutrino state may evolve differently and then the
lepton number violating oscillation may take place. 

The study of neutrino has a lot of cosmological and astrophysical
implications.  For example, neutrinos are the
important ingredient of the energy produced and removed from the center of Sun,
they carry off the largest amount energy of an exploding star in supernova,
neutrinos in certain accretion disks around a compact object are responsible for
cooling process,  
Moreover, neutrinos may be considered as prime dark matter candidate.
Although the mass of a neutrino is very less, because of their large number,
density is very great
and they would thus exert a very potent gravitational effect helping to determine the
rate and the pattern of galaxy formation. Therefore, they may have significant role
in the large scale structure of the universe. Finally, 
the neutrino asymmetry in
early universe may be responsible for leptogenesis and then baryogenesis, etcetera.

In the flat space, the neutrino oscillation is due to difference in rest masses
between two mass eigenstates. However,
Gasperini \cite{gas} first pointed out that presence of gravitational field
affects different neutrino flavors differently which violates equivalence
principle and
thus governs oscillation, even if neutrinos are massless or of degenerate mass.
Subsequently, the effect of strong gravitational field on neutrino oscillation was discussed \cite{hl}.
Later it was shown that the neutrino oscillation 
with LSND data \cite{mansar} can be explained by
degenerate or massless neutrinos with flavor non-diagonal gravitational coupling.
It was further argued \cite{ab} that in weak gravitational
field the flavor oscillation is possible with the 
probability phase proportional to the gravitomagnetic field. 
The oscillation was also shown to be feasible 
when the maximum
velocities of different neutrino differ each other, even if they
are massless \cite{cg}. 

All the above results are for flavor oscillation
or/and without any rigorous effects of
general relativity. However, as we show below explicitly that the nature
of curvature plays very important role to determine whether the oscillation
occurs or not, even under strong gravity. 
Therefore, the goal of the present 
paper is to address the {\it neutrino$-$antineutrino oscillation},
which violates lepton number conservation, 
focusing on the nature of space-time curvature and its special effect. 
In this case, any oscillation depends on the gravitational coupling strength. 
The main message is something new and unique compared to that 
in earlier studies mentioned above and 
many other under curved space-time \cite{all}. 

While the neutrino$-$antineutrino oscillation
under gravity is an interesting issue on its own right, the present 
result is able to address two long-standing
mysteries in astrophysics and cosmology as well as related 
nuclear physics: (1) Source of abnormally large 
neutron abundance to support the r-process nucleosynthesis 
in astrophysical site. (2) Possible origin of baryogenesis.

It has recently been
shown that under gravity, which may bring the CPT violating interaction, energy level
for the neutrino is split up from that of the
antineutrino \cite{sm,m05}. Similar interactions 
not preserving Lorentz and CPT symmetry
were noticed earlier \cite{ck} in different contexts. 
Now the splitting of energy level for neutrinos from that of
corresponding anti-neutrinos may create a difference in their evolutionary phases.
This CPT violation effect is responsible for the transition over neutrino mass eigenstates, 
if such states
which are the linear combination of a neutrino and an antineutrino state are possible
to construct, 
in presence of a mechanism to break the lepton number conservation.

The possible violation of local Lorentz symmetry in the
kaon sector of the standard model was studied with both CPT
conserving and non-conserving cases \cite{hms}. It was also pointed
out \cite{dva2} that in a CPT non-conserved case the
neutrino and the antineutrino acquire different mass and thus
there is a possibility of oscillation between them. 
Later, it was shown \cite{baren} in a CPT non-conserved case that in
absence of neutrino decay a neutrino evolves differently from an anti-neutrino
of the same flavor and thus the oscillation 
between them. Recently, it has been argued \cite{ad} that
a linear superposition of two opposite helicity states can be
described as an eigenstate of the de Sitter Casimirs which
actually gives rise to the Majorana state. However, all these 
works neither explain the explicit physical origin of CPT violation nor demonstrate
the dependence of oscillation upon the physically observable quantities.
In the present paper, we
consider realistic situations when neutrinos are propagating
in a strong gravitational field, e.g. the space-time
close to a compact object or in the early universe era, such that
the spin of a neutrino couples with the
spin-connection to the space-time and thus generates a
gravitational interaction even if there is no other interaction
present. 

In the next section, we describe the basic formalism of the problem starting from 
the fermion Lagrangian density and then how does this generate the oscillation
between a neutrino and an antineutrino state. Subsequently, 
we discuss about the oscillation phase and the oscillation length in various natural
space-times in \S\S III,IV. In \S V, we discuss experimental bound on CPT
violation. Finally we summarize the results in \S VI.

\section{Formalism}
\subsection{Fermion Lagrangian density in curved space-time}

We consider a locally flat coordinate system where the gravitational interaction
comes in as an effective interaction. Then
the neutrino Lagrangian density under gravity 
is \cite{sm,m05}
\begin{eqnarray}
{\cal L}=\sqrt{-g}\,\overline{\psi}\left[(i\gamma^a\partial_a-m)+\gamma^a\gamma^5
B_a\right]\psi ={\cal L}_f+{\cal L}_I, ~~~
\label{lagf}
\end{eqnarray}
where
\begin{eqnarray}
B^d=\epsilon^{abcd} e_{b\lambda}\left(\partial_a e^\lambda_c+\Gamma^\lambda_{\alpha\mu}
e^\alpha_c e^\mu_a\right),
\,\,\,\, e^\alpha_a\, e^\beta_b\,\eta^{ab}=g^{\alpha\beta},
\label{bd}
\end{eqnarray}
where the choice of unit is, $c=\ch=k_B=1$. 
Naturally, the Lagrangian density in a local flat space-time can be split out
into two parts. One is the free part similar to the flat space Lagrangian density 
apart from a multiplicative factor $\sqrt{g}$, and other is the gravitational interaction
part which comes in due to an effective extension
of a flat-space description. 

\subsection{CPT status of the Lagrangian}

${\cal L}_I$ may be a CPT and Lorentz violating interaction if the background
curvature coupling, $B_a$, is constant in the local frame or CPT even \cite{m05}. 
Earlier, similar interaction terms were also
considered in CPT violating theories and string theory without any explicit possible
origin in nature \cite{ck,kp}. 
If  $B_a$ does not flip in
sign under CPT transformation, then ${\cal L}_I$ violates CPT. 
Actually, as described in detail in our previous work \cite{dmd}, 
under the CPT transformation,
associated axial-vector or pseudo-vector ($\bar{\psi}\gamma^a\gamma^5\psi$) changes sign.
If $B_a$ is treated as a background field in a local
frame, then the interaction violates CPT explicitly. 
However, in the present case, with its functional form we can determine
the explicit CPT status of $B_a$ itself at a space-time point $(t,x,y,z)$.
If $B_a(-x,-y,-z,-t)= - B_a(x,y,z,t)$, then $B_a$ is CPT odd. On the other
hand, CPT even $B_a$ corresponds to $B_a(-x,-y,-z,-t)= B_a(x,y,z,t)$.
If $B_a(x,y,z,t)$ is not an odd function under CPT
[$B_a(-x,-y,-z,-t)\neq - B_a(x,y,z,t)$], then ${\cal L}_I$ comes out to be
a CPT violating interaction along the space-time.
It is the nature of background
metric which determines whether $B_a$ is odd or even under CPT and then
the overall CPT status of the interaction \footnote{The
associated axial-vector, $\overline{\psi}\gamma^a\gamma^5\psi$,
always changes sign under CPT.}.

It is very important to note, and can be easily verified, 
that the gravitational coupling, $B_a$, is zero when
the space-time is spherically symmetric and the metric does not consist of any off-diagonal
element, e.g. for the Robertson-Walker universe, the Schwarzschild black hole, etcetera.
Nevertheless, there may exist the spin-orbit coupling 
when the helicity of a neutrino is different from its chirality. However, the 
spin-orbit coupling effect is
expected to be small compared to the effect due to $B_a$ arised in a non-spherical 
symmetric space-time.

\subsection{Energy dispersion relations for neutrinos and anti-neutrinos}

Now we stick to the standard model 
so that a neutrino is solely left-handed and an antineutrino is
solely right-handed. 
Therefore, when the background gravitational field is
constant or $B_a$ is CPT even, 
 ${\cal L}_I$, in Weyl's representation, can be written
explicitly in combination of the neutrino and the antineutrino fields as
\begin{eqnarray}
\overline{\psi}\gamma^a \gamma^5 \psi B_a= \left(\overline{\psi}_R \gamma^a \psi_R
-\overline{\psi}_L \gamma^a \psi_L\right)B_a.
\label{part}
\end{eqnarray}
Then, the dispersion energy relations 
for the neutrino and the anti-neutrino 
are \cite{m05}
\begin{eqnarray}
E_{\nu} =  \sqrt{({\vec p} - {\vec B})^2 + m^2} + B_0, ~~~~
E_{\overline{\nu}} = \sqrt{({\vec p} + {\vec B})^2 + m^2} -
B_0. \label{edis}
\end{eqnarray}
As $E_\nu\neq E_{\overline \nu}$ \footnote{This is clearly due to
presence of background curvature which appears as gravitational
four vector $B_a$.}, the time evolution of a neutrino is different
compared to that of an antineutrino.

Let us now consider the Majorana neutrino when 
the anti-particle is basically
the charged conjugated particle, and thus
one can construct the
Majorana spinor 
which has only left-handed component such that $\psi=\psi_L$ and 
$\psi^c=\psi_L^c=\psi_R$,
which gives rise to the lepton number violating mass terms in ${\cal L}$.  
Therefore, eqn. (\ref{lagf}) can be re-written explicitly for Majorana neutrinos as
\begin{eqnarray}
{\cal L}=\sqrt{-g}\left[\left(i\overline{\psi}_L\gamma^a\partial_a\psi_L+
i\overline{\psi}_L^c\gamma^a\partial_a\psi_L^c\right)
-m\left(\overline{\psi}_L^c\psi_L+\overline{\psi}_L\psi_L^c\right)
+\left(\overline{\psi}_L^c\gamma^a\psi_L^c-\overline{\psi}_L\gamma^a\psi_L\right)B_a\right].
\label{neulagfin}
\end{eqnarray}
Thus, the Lagrangian may break the lepton number conservation and CPT both simultaneously, depending 
of the background, which may lead to oscillation.

\subsection{Oscillation between neutrino and antineutrino states}

Now motivated by the neutral kaon system, we consider two distinct orthonormal 
eigenstates $|E_\nu>$ and $|E_{\overline \nu}>$, for a neutrino and an antineutrino type respectively, 
of the same flavor. 
Further we introduce a set of neutrino mass
eigenstates at $t=0$ as
\begin{eqnarray}
|m_1>=cos\theta\, |E_\nu>+sin\theta\, |E_{\overline \nu}>,\hskip0.5cm
|m_2>=-sin\theta\, |E_\nu>+cos\theta\, |E_{\overline \nu}>.
\label{fl2}
\end{eqnarray}
Therefore, 
the oscillation probability for $|m_1(t)>$ at $t=0$
to $|m_2(t)>$ at a later time $t=t_1$ can be found as
\begin{eqnarray}
\nonumber
P_{12}&=&\left|\left[-sin\theta\, <E_\nu|+cos\theta\, <E_{\overline \nu}|~\right]\left[e^{-iE_\nu t_1}
cos\theta\,|E_\nu>+e^{-iE_{\overline \nu} t_1}sin\theta\,|E_{\overline \nu}>\right]\right|^2\\
&=&sin^22\theta\,sin^2\delta,
\label{pab}
\end{eqnarray}
where for ultra-relativistic neutrinos,
\begin{eqnarray}
\delta=\frac{(E_\nu-E_{\overline \nu})t_1}{2}=
\left[(B_0-|\vec{B}|)+\frac{\Delta m^2}{2|{\vec p}|}\right]\,t_1.
\label{phf}
\end{eqnarray}
The term within parenthesis in the right hand side of eqn. (\ref{phf})
is arised due to gravity, while the other term is the flat space contribution
which is zero when rest masses of the neutrino and the antineutrino are same and/or
in the massless limit. 
In rest of the description we set $\Delta m^2=0$, as our aim is to establish any gravity effect
which is expected to contribute dominantly.

Therefore, a non-zero oscillation phase, due to $B_a\neq 0$, between a neutrino and an anti-neutrino 
is expected to appear which is gravitational in nature. 
The interesting point to note is that the present mechanism
explicitly affects the neutrino$-$antineutrino pair unlike the cases considered
earlier (e.g. \cite{gas,all}) when the gravity would mainly affect 
different flavors.  Note that a general field theoretic description
of neutrino oscillation including the possible neutrino-antineutrino oscillation
has already been worked out and has been shown that CPT violation provides
oscillations without the additional sterile neutrinos \cite{km04}. 
In a similar fashion,
our mechanism also explains neutrino-antineutrino oscillation 
without incorporating sterile neutrinos. In general, it is the non-zero
coefficient for CPT violation in the effective Lagrangian which generates
neutrino-antineutrino oscillation in both the models, while we explicitly
show that the CPT violation coefficient may originate from the background
curvature. Both the models mainly emphasize the possibility of oscillation
in absence of neutrino mass difference. 
The unconventional energy dependence in oscillation 
phase, $\delta$, (for massless neutrinos) given by eqn. (\ref{phf})
also reflects its similarity to the earlier work \cite{km04}.


\section{Oscillation phase in specific space-times}

It is very clear from eqns. (\ref{lagf}), (\ref{edis}) and (\ref{neulagfin}) that due to 
gravity the neutrino 
and the anti-neutrino acquire different effective mass which
finally leads to a non-zero oscillation probability in two
mass eigenstates. 
Therefore, in curved backgrounds, e.g. in the anisotropic phase 
of early universe when the space-time is non-flat, close to a rotating compact object, 
the neutrino$-$antineutrino oscillation exists
in presence of a suitable mechanism to violate lepton number conservation.
Below we describe the oscillation according to specific space-times.

\subsection{Early Universe}

First we consider the anisotropic phase of axially symmetric early universe, when
the GUT processes lead to lepton number non-conservation.
With a simplified version of the Bianchi II model, 
\begin{equation}
d s^2 = -dt^2+S(t)^2\, dx^2+R(t)^2\,[dy^2+f(y)^2\,dz^2]-S(t)^2\,h(y)\,[2dx-h(y)\,dz]\,dz,
\label{mat}
\end{equation}
where $f(y)=y$ and $h(y)=-y^2/2$.
The corresponding orthogonal set of non-vanishing tetrad (vierbien) components can be chosen as
\begin{eqnarray}
\nonumber
&&e^0_t = 1,\,\,e^1_x=f(y)R(t)S(t)/\sqrt{f(y)^2R(t)^2+S(t)^2h(y)^2},\,\,e^2_y=R(t),\\
&&e^3_z=\sqrt{f(y)^2R(t)^2+S(t)^2h(y)^2},\,\,e^3_x=-S(t)^2h(y)/\sqrt{f(y)^2R(t)^2+S(t)^2h(y)^2}.
\label{tetnonv}
\end{eqnarray}
We then obtain the components of $B^d$ 
\begin{eqnarray}
\nonumber
&&B^0=\frac{4R^3S+3y^2R\,S^3-2y\,S^4}{8R^4+2y^2R^2S^2}, \,\,
B^1=0,\\
&&B^2=\frac{(4y\,R^2-8R\,S-y^3S^2)(R\,S^\prime-R^\prime S)}{8R^4+2y^2R^2S^2},\,\,
B^3=0.
\label{bian2}
\end{eqnarray}
It is very clear from above that $B^0$ and $B^2$
do not flip sign under space-inversion,
i.e. for $y\rightarrow -y$. Thus, it is not an odd function over the space-time for
any of the Bianchi models and the form of $B_a$ is such that
$B_a(-x,-y,-z,-t) \neq \pm B_a(x,y,z,t)$. 
An example of a space-time where $B_0(-x,-y,-z,-t)=-B_0(x,y,z,t)$ can be found in 
\cite{mmp} such that ${\cal L}_I$ under CPT does not change sign overall.
Therefore, according to the discussion 
in \S II.B, $B_a$ leads to CPT violation
at any space-time point $(x,y,z,t)$.
Hence, one can obtain the oscillation phase $\delta$ from eqn. (\ref{phf}).
If we consider a special case when 
$S(t)={\rm arbitrary\,\, constant}=k$, then $B^0,B^2\rightarrow 0$ at 
$t\rightarrow\infty$ \footnote{It is easy to compute that $R(t)=(A\,t-B)^{1/2}$, when 
$A$ and $B$ are arbitrary constants \cite{bsc}.}, which verifies that the curvature coupling
is insignificant in the present universe to exhibit any such oscillation.
It also can easily be checked that in a spherically symmetric case, say for 
Robertson-Walker universe, $B^d$ vanishes at all $t$ and thus 
the oscillation probability is zero.

\subsection{Around black holes}

Second example can be given for a space-time around the compact object 
when the curvature effect is significant. 
In early, the flavor oscillations around rotating black holes (in case of active galactic nuclei) were studied
\cite{wudka}. 
The black hole space-time, namely the Kerr geometry, can be described 
in Cartesian-typed coordinate $(t,x,y,z)$ \cite{doran} as
\begin{equation}
d s^2 = \eta_{ij} \, d x^i \, d x^j - \bigg[ \frac{2 \alpha}{\rho} \, s_i \, v_j + \alpha^2 \, v_i \, v_j \bigg] d x^i \, d x^j \label{metbh}
\end{equation}
where
\begin{eqnarray}
\nonumber
&&\alpha = \frac{\sqrt{2 M r}}{\rho}, \, \rho^2 = r^2 + \frac{a^2 z^2}{r^2}, \\
&&v_i = \left(1, ~ \frac{a y}{a^2 + r^2}, ~ \frac{- a x}{a^2 + r^2}, ~ 0 \right),\,
s_i = \left(0, ~ \frac{r x}{\sqrt{r^2 + a^2}}, ~ \frac{r y}{\sqrt{r^2 + a^2}},~ 
\frac{z \sqrt{r^2 + a^2}}{r} \right).
\end{eqnarray}
Here $a$ and $M$ are respectively specific angular momentum and mass of the Kerr
black hole and $r$ is positive definite satisfying,
\begin{equation}
r^4 - r^2 \, \left(x^2 + y^2 + z^2 - a^2 \right) - a^2 z^2 = 0.
\end{equation}
The corresponding non-vanishing components of tetrad (vierbien) are given as \cite{doran}
\begin{eqnarray}
\nonumber
&&e^0_t = 1, ~~ ~~  e^1_t = - \frac{\alpha}{\rho} \, s_1, ~~ ~~ e^2_t = - \frac{\alpha}{\rho} \, s_2, ~~ ~~ e^3_t = - \frac{\alpha}{\rho} \, s_3, \label{tet1}  \\
\nonumber
&&e^1_x = 1 - \frac{\alpha}{\rho} \, s_1 \, v_1, ~~ ~~ e^2_x = - \frac{\alpha}{\rho} s_2 \, v_1, ~~ ~~ e^3_x = - \frac{\alpha}{\rho} \,
s_3 \, v_1, \label{tet2} \\
&&e^1_y = - \frac{\alpha}{\rho} \, s_1 v_2, ~~ ~~ e^2_y = 1 - \frac{\alpha}{\rho} \, s_2 \, v_2, ~~ ~~ e^3_y = - \frac{\alpha}{\rho} \,
s_3 \, v_2, ~~ ~~ e^3_z = 1 - \frac{\alpha}{\rho} \, s_3 \, v_3 \label{tet3}.
\label{tetnonv}
\end{eqnarray}
From eqn. (\ref{bd}) we obtain the gravitational scalar potential
\begin{eqnarray}
 B^0  
&=&e_{1\lambda}  \left( \partial_3 e_2^\lambda - \partial_2 e_3^\lambda
 \right) + e_{2 \lambda} \left( \partial_1 e_3^\lambda - \partial_3 e_1^\lambda
 \right) + e_{3 \lambda} \left( \partial_2 e_1^\lambda - \partial_1 e_2^\lambda \right) 
=-\frac{4a\sqrt{M}z}{\bar{\rho}^2\sqrt{2r^3}},
\label{b0z}
\end{eqnarray}
where $\bar{\rho}^2=2r^2+a^2-x^2-y^2-z^2$.\\
Similarly, one can obtain $B^1,B^2,B^3$.
Clearly, from eqn. (\ref{b0z}),
$B^0(-t,-x,-y,-z)=B^0(t,x,y,z)$; above $B^0$ is a CPT even function which
makes ${\cal L}_I$ a CPT odd interaction.
Therefore, the neutrino-antineutrino oscillation is feasible and the
oscillation phase, $\delta$, can easily be computed.
It is then a trivial job to check that the 
oscillation phase for a non-rotating (Schwarzschild) compact object is zero,
as there $B^d$ itself vanishes,
by setting $a=0$ in the expression of $B^d$.

\section{Oscillation length and its consequence }

The above discussion establishes the fact that the gravity induced neutrino$-$antineutrino 
oscillation is expected to occur
provided the metric consists of 
one off-diagonal term at least. 
The amplitude of oscillation is zero at $\theta=0,\pi/2$ and maximum at
$\theta=\pi/4$. From eqns. (\ref{pab}) and (\ref{phf})  the oscillation length, $L_{osc}$, 
by appropriately setting dimensions, is
obtained as
\begin{eqnarray}
L_{osc}= c\,t_1=\frac{\pi\,\ch\,c}{\tilde{B}}\sim\frac{6.3\times 10^{-19}GeV}{\tilde{B}}{\rm km},
\label{old1}
\end{eqnarray}
where $\tilde{B}=B_0-|\vec{B}|$ is expressed in GeV unit and 
the neutrino is moving in the speed of light.

\subsection{Around black holes}

One of the most fruitful situations for an oscillation of this kind to take place 
is the inner region of an accretion disk around a rotating compact object. 
Therefore, for a compact object of mass $M=M_s\,M_\odot$, the oscillation length at an orbit,
$r\sim\bar{\rho}= x\,M$ and $z=H\,M$, of the disk, assuming $\tilde{B}\sim B_0$, is computed 
from eqns. (\ref{b0z}) and (\ref{old1})
\begin{eqnarray}
L_{osc}\sim \frac{1.8\,x^{7/2}\,M_s}{a\,H}{\rm km}=\frac{1.2\,x^{7/2}}{a\,H}M.
\label{diskos}
\end{eqnarray}
This result has a very important implication to the neutrino dominated
accretion flow (NDAF) \cite{ndaf,knp}. Although the NDAF and the related supernova
are described in the existing literature ignoring any gravity induced 
oscillation effects, it is expected that the oscillation at an
inner edge of the disk to be influenced by gravity that affects the neutrino 
sphere and then the accretion dynamics and outflow which are directly related to the 
corresponding prediction of supernova explosion. It appears from eqn. (\ref{diskos}) that
at the inner accretion 
disk, when $x\le 10$, $L_{osc}$ varies from a few factors to several hundreds of 
Schwarzschild radii, depending on the location and the thickness of the 
disk, for a fast spinning
compact object. This is interesting as the accretion 
disk can be extended upto several thousands of
Schwarzschild radii.

One of its important consequences is to the r-process nucleosynthesis.
Supernova is thought to be the astrophysical site of the r-process nucleosynthesis. 
During supernova, neutron capture processes for radioactive elements take place
in presence of abnormally large neutron flux. However, how does the large neutron flux
arise is still an open question. There are two related reactions:
\begin{equation}
\,\,n+\nu_e\rightarrow p+e^-,\,\,\,\, \,\,\,\,\,p+\bar{\nu}_e\rightarrow n+e^+.
\label{nuc}
\end{equation}
If $\bar{\nu}_e$ is over abundant than $\nu_e$, then from eqn. (\ref{nuc}) neutron production
is expected to be more than proton production into the system. Therefore, the possible
conversion of $\nu_e$ to $\bar{\nu}_e$ due to the gravity induced oscillation 
explains the overabundance of neutron.

\subsection{Early Universe}

If the oscillation is considered in early universe with anisotropic phase, then
at the GUT scale $\tilde{B}\sim 10^5$ GeV 
\cite{dmd}. From eqn. (\ref{old1}), this leads to 
$L_{osc}\sim 10^{-24}$km which is $10^{14}$ orders of magnitude larger than the
Planck length. This has an important implication as the size 
of universe, $\chi=\int_0^tcdt^\prime/R(t^\prime)$,
at the GUT era is within $\sim 10^{26}$ times of the Planck. Therefore, the oscillation
may lead to leptogenesis and then to baryogenesis by electro-weak sphaleron processes
due to $B-L$ conservation, what we see today. 

\subsection{Atmosphere of Earth}

If we consider a case of neutrino$-$antineutrino oscillation 
in the atmosphere of earth where the curvature
effect, $\tilde{B}\sim 10^{-37}$ GeV, can be found on a satellite orbiting earth with
velocity $v_\phi\sim 1$ km/sec \cite{mmp}, then $L_{osc}$ at that
orbit comes out to be of the order of $10^{18}$ km. This is quite large that 
the satellite revolves about $10^{13}$ times in the time taken to complete only
one oscillation. This
is because the neutrino$-$antineutrino oscillation is difficult to observe in 
the atmosphere of earth.

\section{Bounds on CPT violation from experiments and our model}

Kosteleck\'y and his collaborators \cite{ck} already argued that the CPT violation, 
if present, would be a very small effect and then they argued for very stringent 
constraints to the corresponding coefficients. The experiment E773 at Fermilab
published a bound to mass difference between particle and antiparticle in
the neutral kaon sector as
$r_K=|m_{K^0}-m_{\bar {K^0}}|/m_{K^0} < 1.3\times 10^{-18}$ \cite{fermi}. 
It was also argued that any observable effects must be suppressed due to
involvement of, perhaps, higher-level fields of Planck scale mass 
$M_{Pl}$ \cite{kpot}. When
the electroweak scale mass $m_{ew}\sim 10^2$GeV, $m_{ew}/M_{Pl}\lsim 10^{-17}$ 
provides the natural dimensionless quantity which governs the suppression.
In the neutral kaon system, one could expect
$r_K\sim m_{ew}/M_{Pl}$ which is just below the present bound.
The same authors also pointed out that $m_{ew}/M_{Pl}$ might be expected
to arise in a gravitational mechanism that violates CPT.

In the present case, we have constructed the neutrino mass eigenstates 
under gravity described
in \S II.D in the spirit of the neutral kaon mass matrix when the
mass difference between neutrino and antineutrino is $\sim 2\tilde{B}$. Now from \S IV,
for black holes of mass range, e.g. $10M_\odot\le M_s\le 10^6M_\odot$, the
oscillation length comes out to be $10{\rm km}\lsim L_{osc}\lsim 10^7$km, 
(which is equivalent to a few factors to a few hundreds Schwarzschild radii,
as mentioned in \S IV.A), 
provided $10^{-19}{\rm GeV} \gsim \tilde{B}\gsim 10^{-28}$GeV. Therefore, 
the necessary $\tilde{B}$ to govern any significant oscillation in an
accretion disk is very small and is quite compatible with the experimental bound.

In the GUT scale, as described in \S IV.B, the physically interesting
oscillation length may range e.g. as
$10^{-24}{\rm km}\lsim L_{osc}\lsim 10^{-9}$km, provided
$10^{5}{\rm GeV} \gsim \tilde{B}\gsim 10^{-10}$GeV, when size of the 
universe is $\lsim 10^{-12}$km. 
If we compare this $\tilde{B}$ with the mass scale ($m_{\rm GUT}\sim 10^{16}$GeV)
of the system, then it comes out to be one part in $10^{11}$ to one part in $10^{26}$.  

Much more stringent bounds can be put on the gravitational pseudo four-vector 
potential $B_a$
(which is equivalent to $\tilde{B}$) from the tests for CPT and Lorentz violation
\cite{kbook}. ${\cal L}_I$ in non-relativistic limit is equivalent to 
$\vec{s}.\vec{B}$ due to
the interaction between the fermion spin $\vec{s}$ and the external field $\vec{B}$.
One can measure this interaction energy in experiments where a macroscopic 
number of fermions can be polarized in the same direction. For example, in the
Eot-Wash II experiment \cite{eot}, the spin-polarized torsion balance has 
$N=8\times10^{22}$ aligned spins (with very negligible net magnetic moment).
There will be a torque $\tau=(N/\pi)\Delta E$ on such a torsion balance 
where $\Delta E=|\vec{B}|$ is the difference in energy between the fermion 
spins polarized parallel and antiparallel to the external $\vec{B}$ field.
From this experiment \cite{eot}, one can measure up to $|\vec{B}|\sim 10^{-28}$GeV
\cite{bk}. The field $\vec{B}$ also can be probed by measuring the net magnetization
in a paramagnetic material using a squid \cite{ni}. An external field $\vec{B}$ 
appears as an effective magnetic field of strength $\vec{B}_{\rm eff}=(\vec{B}/\mu_B)$.
The magnitude of $\vec{B}_{\rm eff}$ can be probed in this experiment as 
$|\vec{B}_{\rm eff}|=10^{-12}$G which reflects $\vec{B}\sim 10^{-29}$GeV \cite{bk,ni}.

\section{Summary}

Although the flavor neutrino oscillation has already
been described in several occasions successfully,
the underlying mechanism for the neutrino$-$antineutrino oscillation is still not 
well understood. While in the first case, lepton number of the system remains
conserved, the second case is a lepton number violating phenomenon that causes
it more difficult to explain and establish. We have argued that in presence
of space-time curvature violating CPT symmetry, the energy level of a neutrino
splits up from that of an antineutrino which gives rise to their different 
effective mass. This affects their distribution and
results their difference in phases of evolution.
Then, motivated by the neutral kaon
system, we have described two mass eigenstates as linear combination
of a neutrino typed and an antineutrino typed state.
Therefore, the oscillation between mass eigenstates is expected.

We have explicitly demonstrated this lepton number violating 
oscillation with some natural examples. We have shown that in the
anisotropic phase of early Universe, when the space-time is non-flat,
the neutrino$-$antineutrino oscillation is feasible which may lead
to the leptogenesis and then to the baryogenesis in presence of a suitable
mechanism to violate the lepton number conservation. This may be
the possible origin of the matter-antimatter asymmetry in Universe 
what we see today.

On the other hand, in the inner region of an NDAF
where the general relativistic effect is important, this
oscillation may affect the accretion dynamics and outflow.
It has recently been argued \cite{knp} that the energy carried
out in the outflowing wind from such disks could be as high as $10^{51}$ erg
which might be sufficient to convert a failed supernova explosion into a 
successful one. They have also noted that the neutron to proton ratio is
large in several regions of the disk. The neutron-rich regions are
thought to be the astrophysical site for r-process nucleosynthesis.
However, it is unclear that how does the large neutron abundance over
proton arise in the first place. Our calculation suggests that
this is due to larger antineutrino, due to oscillation, 
over the neutrino which favors
the proton to decay into neutron compared to the neutron to proton.
One now needs to study the NDAF in detail incorporating oscillation
effects and to see how does the disk structure including neutrino sphere 
change and then affect the energy budget.

\vskip0.8cm
\noindent{\large Acknowledgment}\\
The author acknowledges the partial support to this research by
NSF grant AST 0307433.

\end{document}